\documentclass{paper}
\usepackage{epsfig, graphicx}
\usepackage[english]{babel}

\begin{document}
\title{Nonlinear Absorption of Radio Waves in a Noncollinear Antiferromagnet}
\author{\small Aleksey M. Tikhonov\/\thanks{tikhonov@kapitza.ras.ru} and Nikolay G. Pavlov}
\maketitle
\leftline{\it Kapitza Institute for Physical Problems, Russian Academy of Sciences,}
\leftline{\it ul. Kosygina 2, Moscow, 119334, Russia}

\rightline{\today}

\abstract{The nonlinear absorption of radio waves (${200 - 800}$\,MHz) in a noncollinear cubic antiferromagnet Mn$_3$Al$_2$Ge$_3$O$_{12}$ in an external magnetic field ${{\bf H}\parallel[001]}$ has been studied in the temperature range of $1.2 - 4.2$\,K. We attribute the observed dissipation of the electromagnetic energy to the parametric excitation of inhomogeneous surface waves at the boundaries of antiferromagnetic domains.}

\vspace{0.25in}

\large

\begin{figure}
\hspace{0.75in}
\epsfig{file=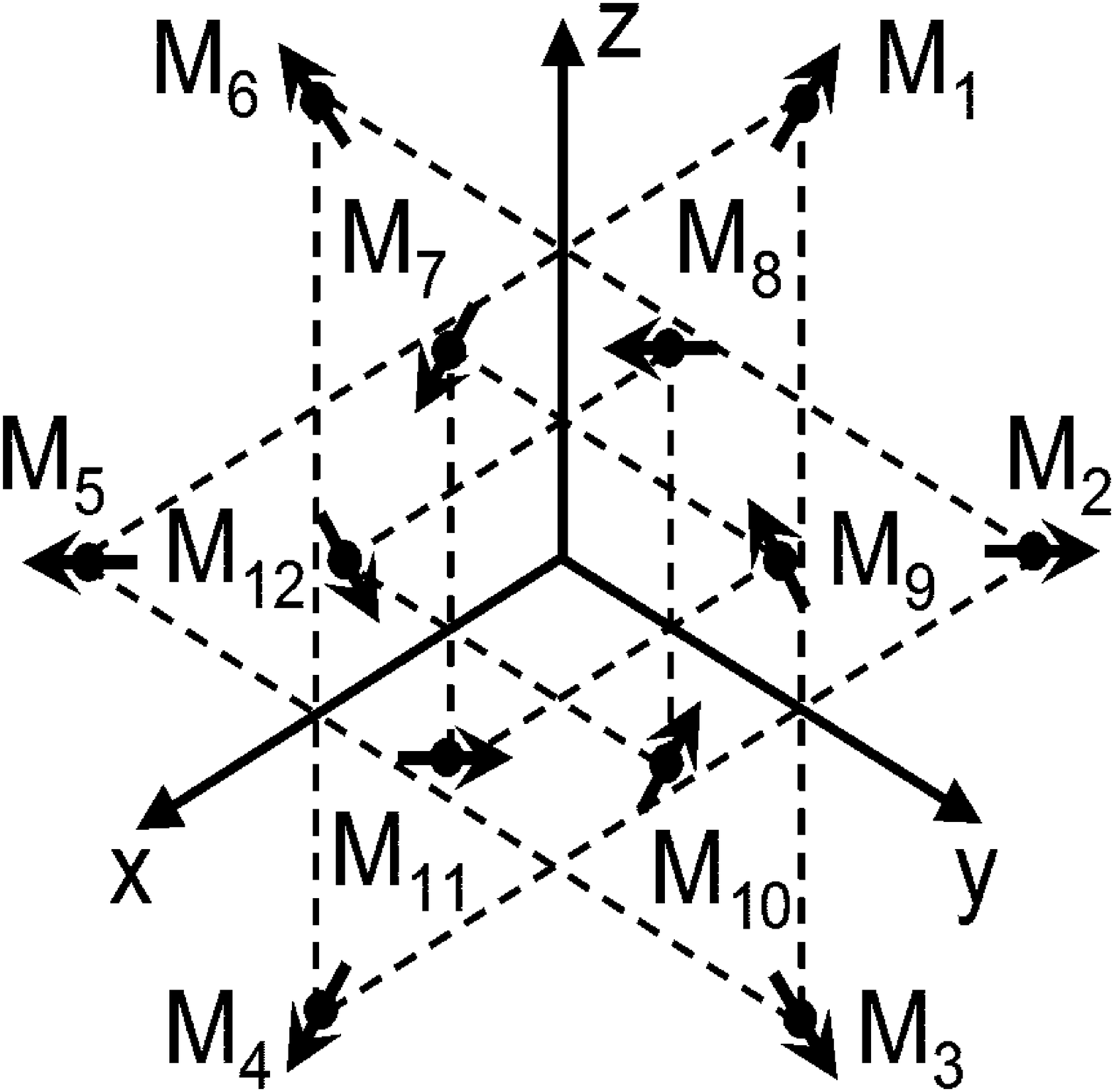, width=0.6 \textwidth}

Figure 1.  Magnetic structure of noncollinear Mn$_3$Al$_2$Ge$_3$O$_{12}$ antiferromagnet in the exchange approximation.
\end{figure}

Owing to the multiple degeneration of the ground
state of a noncollinear cubic antiferromagnet
Mn$_3$Al$_2$Ge$_3$O$_{12}$, a multidomain structure stable in a
wide range of magnetic fields can exist in it. In this
work, we detect the nonlinear absorption of ultrashort
radio waves in the manganese garnet Mn$_3$Al$_2$Ge$_3$O$_{12}$
and attribute it to the parametric excitation of inhomogeneous oscillations of the boundaries of antiferromagnetic domains.

Garnet transits to an antiferromagnetic state at a temperature of about 6.8 K \cite{1}. According to neutron-diffraction studies, a planar 12-sublattice noncollinear
structure (crystal symmetry group O$^{10}_h$) is implemented in it: the magnetic moments of Mn$^{2+}$ are coplanar to the (111) plane and collinear to the [211],
[121], and [112] directions (see Fig. 1) \cite{2,3}. When the external magnetic field ${\bf H}$ is applied along the [001] direction, the rotation of the spin plane occurs and ends when the external field reaches the critical value ${H_c\approx 2.4}$\,T \cite{4,5}.

In the exchange approximation, the magnetic structure of garnet is described by a pair of 
antiferromagnetic vectors ${\bf l}_1$, è ${\bf l}_2$ (${{\bf l}_1 \perp {\bf l}_2}$ è ${\bf l}^2_1={\bf l}^2_2=1$) \cite{6}. Analysis shows that the ground state is fourfold degenerate in the directions of the vector ${\bf n} = [{\bf l}_1{\bf l}_2]$, which
can be collinear to the $[1\,1\,1]$, $[\overline{1}\,1\,1]$, $[1\,\overline{1}\,1]$, and $[\overline{1}\,\overline{1}\,1]$ directions. In the magnetic field ${{\bf H}\parallel[001]}$, this degeneration holds up to the phase transition field $H_c$, above
which ${{\bf n}\parallel[001]}$.

Thus, four types of antiferromagnetic domains with
different orientations of the vector ${\bf n}$ can coexist in a
garnet crystal. The domain structure holds at a small deviation 
of the field $\bf H$ from the fourth-order axis until the field in the (110) plane is ${\leq 700}$\,Oe \cite{4}.

The magnetic field along the [001] direction with
relative inhomogeneity less than 0.1\% in the size of the
sample was created by a superconducting solenoid.

The single-crystal sample was either directly
immersed in a bath with liquid helium or placed in a
vacuum chamber with a heat-exchange $^4$He gas, which
was in a bath with liquid helium. The temperature $T =
1.2 - 4.2$\,K was controlled in the experiment by the
pressure of saturated helium vapor in the bath. In our
experiments, we used the broadband resonance system
of the "split-ring" type \cite{7}. The transmitted power of
the radio-frequency field ${\bf h}$ was detected by a planar
diode when the magnetic field was varied at a fixed frequency 
$\omega$ (its stability in the experiment was ${\Delta\omega/\omega \sim 10^{-5}}$).

We previously studied the magnetic structure of the
noncollinear antiferromagnet Mn$_3$Al$_2$Ge$_3$O$_{12}$ by analyzing the spectra
of nuclear magnetic resonance (NMR) of $^{55}$Mn in the linear absorption regime 
\cite{5}. In those experiments, three NMR lines were observed in
a narrow frequency range near 30\,MHz in fields $H<H_c$ only when the radio 
frequency field had the polarization ${{\bf h}\parallel{\bf H}}$.

\begin{figure}
\hspace{0.5in}
\epsfig{file=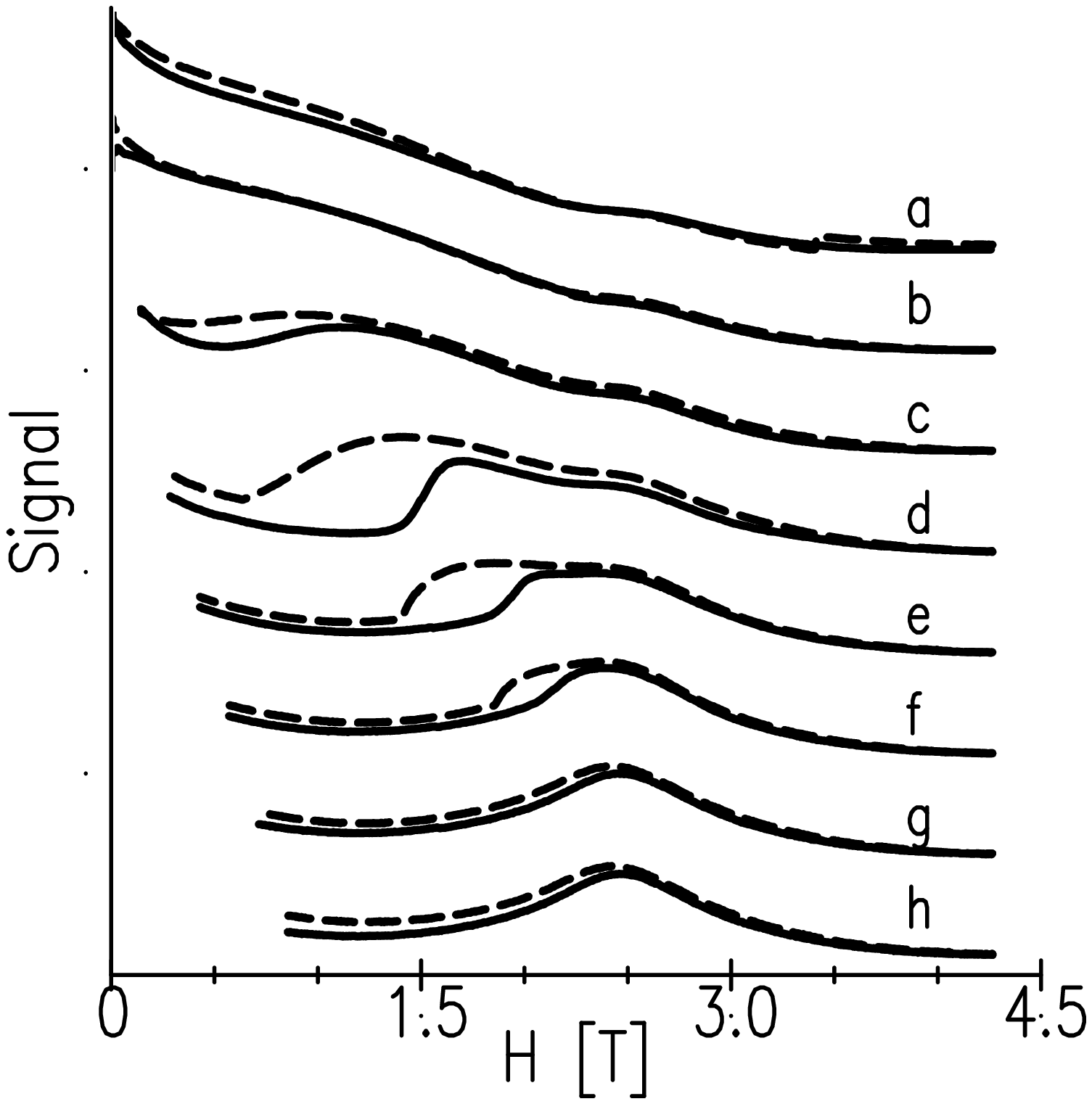, width=0.7 \textwidth}

Figure 2. Power dependencies of the shape of the detector signal
at a frequency of 685\,MHz in Mn$_3$Al$_2$Ge$_3$O$_{12}$ in an
external magnetic field ${{\bf H}\parallel[001]}$ for the polarization of the
radio-frequency field ${{\bf h}\perp {\bf H}}$ and $T=1.2$\,K. The solid and
dashed lines correspond to an increase and a decrease in the magnetic field of the solenoid.
The letters mark the shapes of the signal corresponding to various radio frequency
powers supplied to the resonance system: (a) 20, (b) 19, (c) 18, (d) 17, (e) 16,
(f) 15, (g) 13, and (h) 10\,dBm.
\end{figure}

In this work, we observe the absorption of radio
waves with the polarization ${{\bf h}\perp{\bf H}}$ in the continuous
frequency range of $200 - 800$\,ÌÃö when the amplitude
of the radio frequency field is above a certain threshold
value. Such a spectrum cannot be explained by the resonance properties of the 
nuclear system in Mn$_3$Al$_2$Ge$_3$O$_{12}$ at ${H<H_c}$.

Figure 2 shows the power dependencies of the shape
of the detector signal at a frequency of 685\,MHz in
Mn$_3$Al$_2$Ge$_3$O$_{12}$ for the polarization of the radio-frequency field
${{\bf h}\perp {\bf H}}$ and a temperature of 1.2\,K. The
letters mark the shapes of the signal corresponding to
various radio frequency powers supplied to the resonance system: 
(a) 20, (b) 19, (c) 18, (d) 17, (e) 16, (f) 15, (g) 13, and (h) 10\,dBm 
(in decibels with respect to 1\,mW). The solid and dashed lines correspond to an
increase and a decrease in the magnetic field of the
solenoid. The resonance absorption corresponding to
a branch of low-frequency electron - nuclear oscillations in the 
high-field phase of manganese garnet ($H>H_c$) 
is observed in all curves in a field of about $\sim 2.5$\,T.
When the pump power is above 15\,dBm, the shape of
the signal changes qualitatively: additional absorption
appears at $H<H_c$. In a range from 15 to 18\,dBm, hysteresis 
is also observed in the magnetic field scans.
Such phenomena are not observed for the radio frequency 
field with the polarization ${{\bf h}\parallel{\bf H}}$.

Figure 3 shows the temperature dependencies of the
shape of the detector signal for the polarization ${{\bf h}\perp {\bf H}}$
in the temperature range of $1.2 - 4.2$\,K. The letters
mark the shapes of the signal corresponding to the
temperatures of (a) 4.2, (b) 3.3, (c) 2.0, (d) 1.6, and (e)
1.2\,K. At a temperature of 1.2\,K and a power of
10\,dBm, only resonance absorption is observed near
$H_c$. With an increase in the temperature, an additional
absorption signal appears. At a temperature of 4.2\,K, it
is observed in all fields $H<H_c$. Such qualitative
changes in the signal shape do not occur with an
increase in the temperature at ${{\bf h}\parallel{\bf H}}$.

The parametric excitation of nuclear spin waves in
the bulk of MnCO$_3$ and CsMnF$_3$ crystals was studies in
\cite{8,9} under the conditions of longitudinal radio frequency 
pumping and double resonance with the use of the dependence 
of the position of the antiferromagnetic resonance line on 
the temperature of the nuclear magnetic system. Such phenomena can 
also apparently be observed in manganese garnet, but only in a
narrow range near $H_c$ and at the polarization ${{\bf h}\parallel {\bf H}}$
different from that used in this work. Furthermore, at ${H<0.8H_c}$, 
in view of the features of the spectrum of electron–nuclear oscillations 
in Mn$_3$Al$_2$Ge$_3$O$_{12}$, the frequency range of nuclear magnons 
contracts strongly to the range of $10 - 40$\,MHz near 620\,MHz \cite{5}. 
Thus, the corresponding frequency of the parametric pumping
should be $> 1200$\,MHz, which is noticeably higher than radio frequencies 
in our experiment. The observed phenomenon also cannot be explained by the
resonance creation of acoustic phonons. Indeed, the cross section for this 
process should be independent of $H$. Therefore, absorption would be observed at $H>H_c$.
However, this was not observed in the experiment.

\begin{figure}
\hspace{0.5in}
\epsfig{file=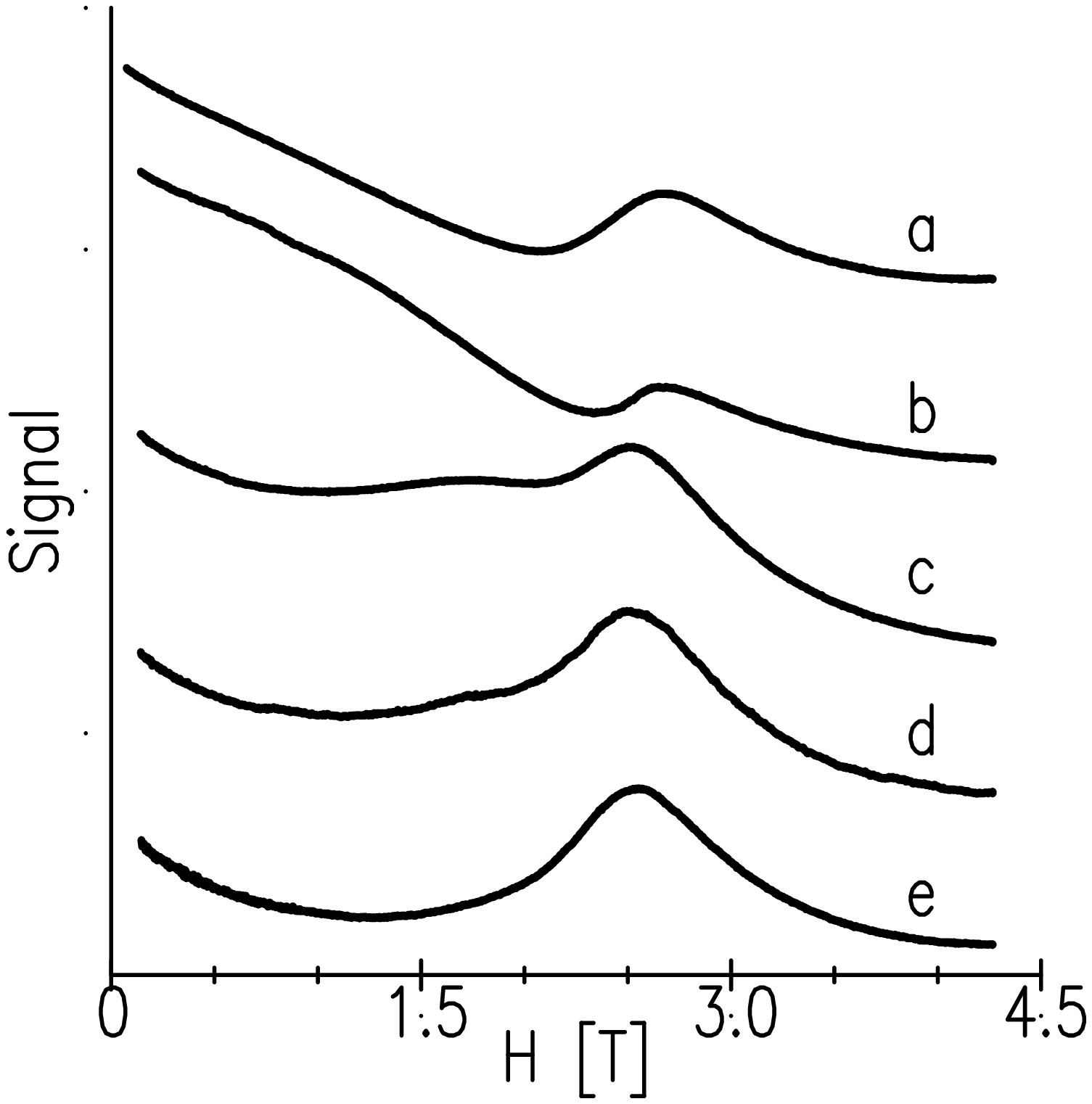, width=0.7\textwidth}

Figure 3. Temperature dependences of the shape of the detector signal at a frequency of
530\,MHz in Mn$_3$Al$_2$Ge$_3$O$_{12}$ at a constant power and ${{\bf h}\perp {\bf H}}$ in the temperature range of $1.2 - 4.2$\,K. The letters mark the shapes of the signal corresponding
to the temperatures of (a) 4.2, (b) 3.3, (c) 2.0, (d) 1.6, and (e) 1.2 K.
\end{figure}

In the first approximation, the radio frequency
field $h(t)$ for the electronic system in Mn$_3$Al$_2$Ge$_3$O$_{12}$
can be considered as quasistatic, because its frequency
($\omega/2\pi< 1$\,GHz) is much lower than the frequencies of
antiferromagnetic resonance whose branches are
above 20\,GHz \cite{6}. The magnetic energy density
depending on the orientation of the magnetic field
${{\bf H}+{\bf h}}$ with respect to ${\bf n}$ has the form
\begin{equation}
\frac{1}{2}\left(\chi_{\perp} -\chi_{\parallel}\right) \left(\bf{H}+\bf{h},\bf{n}\right)^2,
\end{equation}
where $\chi_{\perp}$ and $\chi_{\parallel}$ are the components of the susceptibility 
tensor in the spin plane and parallel to the direction
${\bf n}$, respectively. In our experiment, ${{\bf H}\parallel[001]}$ and the
linearly polarized field ${\bf h}$ lies in the (110) plane. Thus,
the difference between the magnetic energy density of
domains with ${{\bf n}_1\parallel[1\,1\,1]}$ and 
${{\bf n}_2\parallel[\overline{1}\,\overline{1}\,1]}$ is (see Fig. 4)
\begin{equation}
\left(\chi_{\parallel} -\chi_{\perp}\right) Hh(t)\sin{2\alpha},
\end{equation}
where $\alpha$ is the angle between the vectors $\bf H$ and ${\bf n_{1,2}}$.
Expression (2) specifies the force that acts on a unit
area of the boundary between domains with ${\bf n}_1$ and ${\bf n}_2$
and induces its homogeneous oscillations.

\begin{figure}
\hspace{0.5in}
\epsfig{file=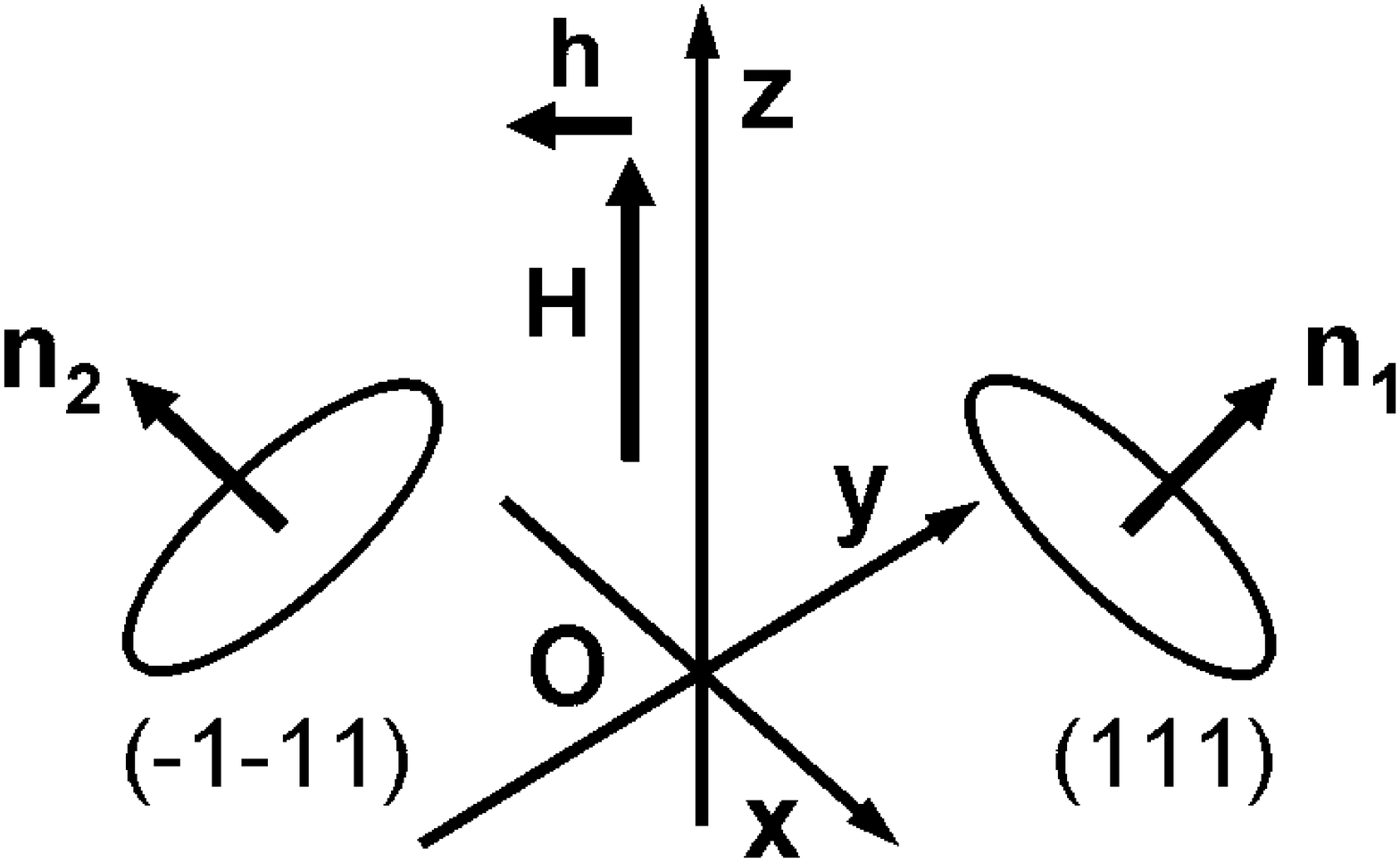, width=0.7\textwidth}

Figure 4. Static magnetic field $\bf{H}$ is oriented along the fourth-
order axis of garnet ([001] direction or the $Oz$ axis),
whereas the alternating field component ${\bf h}$ is applied in the
(110) or $(xy)$ plane. The force ${\left(\chi_{\parallel} -\chi_{\perp}\right) Hh(t)\sin{2\alpha}}$
acts on a unit area of the boundary of domains with ${{\bf n}_1\parallel[1\,1\,1]}$ and
${{\bf n}_2\parallel[\overline{1}\,\overline{1}\,1]}$.

\end{figure}

The anomalous absorption of radio waves can be
considered as the parametric excitation of inhomogeneous oscillations under 
the action of force (2) \cite{10}. If
$\omega_s({\bf k})$ is the frequency of oscillations of the surface
with the wave vector {\bf k}, the absorption of a radio frequency field 
quantum can be accompanied by the resonance creation of two quanta 
of surface waves: $\omega=\omega_s({\bf k})+\omega_s(-{\bf k})$ ( $\hbar\omega_s({\bf k})$ $\approx\hbar\omega/2\sim 10^{-2}$\,K).

We emphasize that Fig. 3 demonstrates a decrease
in the threshold power of the parametric excitation of
domain boundaries with an increase in the temperature $T$. 
With an increase in the temperature, driving force (2) also decreases because of
a decrease in the difference between the susceptibilities $\chi_{\perp}$ and $\chi_{\parallel}$ \cite{4}. Thus, with an increase in the temperature $T$, the effective relaxation in the system decreases sharply.

We are grateful to B.V. Mill for manganese garnet single crystals and 
to Prof. V. I. Marchenko for stimulating discussions and assistance in the experiments.

\vfill\eject

\end{document}